\documentclass[aps,prl,superscriptaddress,amsmath,amssymb,aps,twocolumn,]{revtex4-2}

\usepackage{graphicx}
\usepackage{dcolumn}
\usepackage{bm}
\usepackage{latexsym,epsfig}
\usepackage{graphicx}
\usepackage{verbatim}
\usepackage{comment}
\usepackage{amsmath}
\usepackage{amsfonts}
\usepackage{amsthm}
\usepackage{amssymb}
\usepackage{mathrsfs}
\usepackage{stmaryrd}
\usepackage{color}
\usepackage{physics}
\usepackage{epstopdf}
\usepackage{grffile}
\usepackage{bbold}
\usepackage[normalem]{ulem}
\DeclareGraphicsExtensions{.eps}

\newcommand{\beq}{\begin{equation}}
\newcommand{\eeq}{\end{equation}}
\newcommand{\bea}{\begin{eqnarray}}
\newcommand{\eea}{\end{eqnarray}}
\newcommand{\ben}{\begin{eqnarray*}}
\newcommand{\een}{\end{eqnarray*}}
\newcommand{\bfig}{\begin{figure}}
\newcommand{\efig}{\end{figure}}

\usepackage{hyperref}
\hypersetup{
    colorlinks=true,      
    urlcolor=blue,
    citecolor=blue,
    linkcolor=blue
}

\begin{document}
\title{Floquet realization of prethermal Meissner phase in a two-leg flux ladder}

\author{Biswajit Paul}
\affiliation{School of Physical Sciences, National Institute of Science Education and Research, Jatni 752050, India}
\affiliation{Homi Bhabha National Institute, Training School Complex, Anushaktinagar, Mumbai 400094, India}

\author{Tapan Mishra}
\affiliation{School of Physical Sciences, National Institute of Science Education and Research, Jatni 752050, India}
\affiliation{Homi Bhabha National Institute, Training School Complex, Anushaktinagar, Mumbai 400094, India}

\author{K. Sengupta} 
\affiliation{School of Physical Sciences, Indian Association for the Cultivation of Science, Jadavpur, Kolkata 700032, India.}

\date{\today}

\begin{abstract}
We show that a periodically driven two-leg flux ladder hosting interacting hardcore bosons exhibits a prethermal Meissner phase for large drive amplitudes and at special drive frequencies. Such a prethermal Meissner phase is characterized by a finite time-averaged chiral current. We find an analytic expression of these frequencies using Floquet perturbation theory. Our analysis reveals that the presence of the prethermal Meissner phase is tied to the emergence of strong Hilbert space fragmentation in these driven ladders. We support our analytical results by numerical study of finite-size flux ladders using exact diagonalization and discuss experiments using ultracold dipolar atom platforms that may test our theory. 
\end{abstract}

\maketitle
{\em Introduction:}
Two-leg flux ladders constitute simplest lattice models that can be used to study the effect of external magnetic field on quantum mechanical particles \cite{Giamarchi_lad_2001, Cha_lad_2011,dhar_lad_2012, dhar_lad_2013, dhar_mishra_2013, mishra_lad_2013, wei_lad_2014, Hugel_lad_2014, Tokuno_lad_2014, Kele_lad_2015, Mishra_lad_2016, Rashi_lad_2017, Rashi_lad_2018, jia_lad_2020, ceven_lad_2022, Halati_lad_2023, giri_lad_2023, giri_lad_2024, Huang_lad_2025, pan_lad_2024}. Bosonic ladders with a constant magnetic flux, exhibit Meissner and vortex phases; the former (latter) phases hosts a finite particle current along the boundary (currents along both legs and rungs) of the ladder \cite{Giamarchi_lad_2001, Petrescu_lad_2013, Hugel_lad_2014, Petrescu_lad_2015, Greschner_lad_2015, Greschner_lad1_2015, Piraud_lad_2015, Greschner_lad_2016, Buser_lad_2020, luca_lad_2023,jian_lad_2024}. These systems have been recently experimentally studied using ultracold atoms in optical latices \cite{Ketterle_2013, Monika_2013, Atala2014, Fallani2015, Aidelsburger2015, Gadway_2017, Tai2017, Li2023, impertro2025}. Most of these experiments involving flux-induced phases studied either non-interacting bosons or ladders with a few interacting particles~\cite{Ketterle_2013, Monika_2013, Atala2014, Fallani2015, Aidelsburger2015, Gadway_2017, Tai2017, Li2023}; however, a recent experiment has addressed strong correlation effects in flux ladders~\cite{impertro2025}.

Non-equilibrium dynamics of interacting closed quantum systems have been a subject of intense research in recent years \cite{rev1,rev2,rev3_1,rev3_2,rev4,rev5,rev6, rev7,rev8, rev9, rev10, rev11, rev12}. Periodically driven closed quantum systems have been intensively studied in this context \cite{rev5,rev6, rev7,rev8, rev9, rev10, rev11, rev12}; they exhibit several phenomena such as dynamical freezing \cite{dynfr1_1, dynfr1_2, dynfr1_3,dynfr2_1,dynfr2_2,dynfr2_3,dynfr2_4,dynfr3_1,dynfr3_2,dynfr4_1, dynfr4_2}, dynamical localization \cite{dloc1_1,dloc1_2,dloc2_1,dloc2_2,dloc3_1, dloc3_2,dloc4_1,dloc4_2,dloc5_1,dloc5_2}, emergent topology \cite{topo1_1, topo1_2, topo1_3,topo2_1,topo2_2,topo3_1,topo3_2,topo4_1,topo4_2,topo4_3, Jangjan2020, Wintersperger2020, Nur2023, Nur2025}, realization of Floquet scars \cite{qscar1_1, qscar1_2,qscar2_1,qscar2_2,qscar2_3}, and time-crystalline phase of driven matter \cite{tc1_1, tc1_2,tc2_1, tc2_2, tc2_3,tc3_1,tc3_2,tc3_3,tc4_1,tc4_2,tc4_3} that have no analogue in quantum systems in equilibrium or those subjected to aperiodic drives. One of the key challenges for such driven systems is to control heating which destabilizes possible Floquet phases \cite{rev8}. 
\begin{figure}[t!]
    \centering
    \includegraphics[width=1\columnwidth]{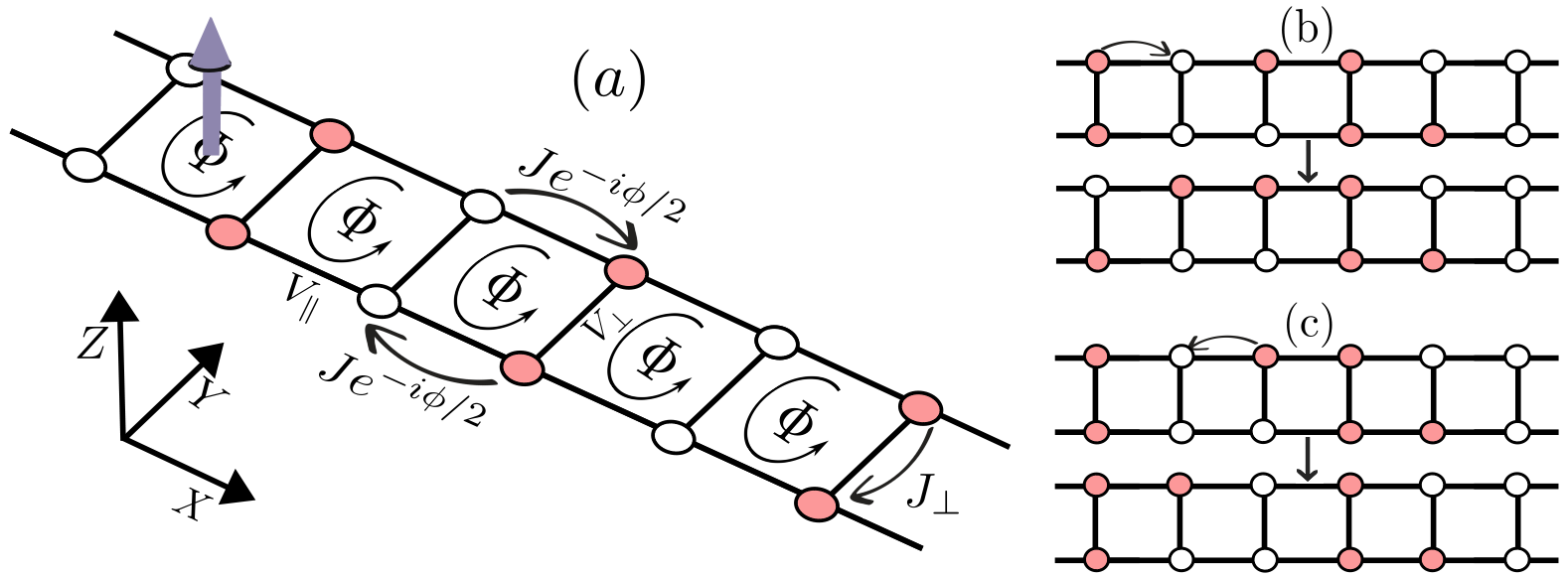}
    \caption{(a) Schematic representation of the two-leg ladder with inter- and intra-leg hoppings (interactions) $J_{\perp}$ and $J$ ($V_{\perp}$ and $V_\parallel$) respectively. The occupied (empty) sites are shown by filled (empty) circles; two filled sites across a rung constitute a {\it bosonic dipole}. The flux $\Phi$ through each plaquette; the corresponding Peierls phase is $\phi= 2\pi \Phi/\Phi_0$. Schematic representation of hopping processes which does (b) and does not (c) change the number of {\it bosonic dipoles}.}
    \label{fig1}
\end{figure}

In this Letter,  we demonstrate Floquet realization of the well-known Meissner phase over a long prethermal timescale in a driven bosonic flux ladder. By considering a system of hardcore bosons interacting via nearest-neighbor (NN) repulsion, schematically shown in Fig.\ \ref{fig1}(a), we  demonstrate that a drive of the inter-leg interaction with large amplitude can lead to the emergence of a prethermal Meissner phase at special drive frequencies. Such a phase is characterized by a finite time-averaged chiral current around the ladder. We show that this emergent Meissner phase can be tied to the presence of an emergent, approximate, Hilbert space fragmentation (HSF) \cite{fraglit1_1,fraglit1_2,fraglit2_1,fraglit2_2,fraglit2_3,fraglit2_4,fraglit2_5,fraglit3_1,fraglit3_2,fraglit3_3,fraglit3_4,fraglit3_5,fraglit4_1,fraglit4_2,fraglit5_1,fraglit5_2,fraglit5_3,flhsf1_1,flhsf1_2,flhsf2_1,flhsf2_2}, of the driven bosonic ladders. We substantiate these facts by exact numerical calculations and perturbative semi-analytical arguments, which allows us to provide an analytic expression of the special frequencies. We also chart out experiments involving ultracold dipolar atoms in optical lattices that can test our theory.

{\em Model:}
\label{model}
We consider hardcore bosons on a two-leg flux ladder with inter- and intra-leg nearest-neighbor interaction
of amplitude $V_{\perp}$ and $V_{\parallel}$ respectively as shown schematically in Fig.\ \ref{fig1} (a). These bosons are  
subjected to periodic driving of $V_{\perp}$ according to square pulse protocol: $V_\perp(t) = V_0$ \text{if $t\le \frac{T}{2}$} and $V_\perp(t) = -V_0$ \text{if $t > \frac{T}{2}$}. Such a system may be modeled by the Hamiltonian $\hat{H(}t) = \hat{H}_0(t) + \hat{H}_1$, where $\hat{H}_0(t) = \sum_{j} V_{\perp}(t)\hat{n}_{j,A}\hat{n}_{j, B}$ and 
\begin{eqnarray}
&& \hat{H}_1 =-J\sum_{j}\left(~e^{-i\phi/2} \hat{b}_{j,A}^{\dagger}\hat{b}_{j+1,A}+e^{i\phi/2} \hat{b}_{j,B}^{\dagger}\hat{b}_{j+1,B}+H.c.\right) \nonumber\\
&&-J_\perp\sum_{j}\left(\hat{b}_{j,A}^{\dagger}\hat{b}_{j,B}+H.c.\right) + V_{\parallel}\sum_{j, \sigma = A, B} \hat{n}_{j, \sigma}\hat{n}_{j+1, \sigma}. \label{eq:Ham}
\end{eqnarray}
Here $\sigma = A, B$ denotes the leg-index, $\hat{b}_{j,\sigma}^{\dagger}$($\hat{b}_{j,\sigma}$) are the bosonic creation (annihilation) operators, and $\hat{n}_{j, \sigma} = \hat{b}_{j,\sigma}^{\dagger}\hat{b}_{j,\sigma}$ is the boson number operator on rung $j$ and leg-$\sigma$. In Eq.\ \ref{eq:Ham}, $J$ and $J_\perp$ denote intra- and inter-leg hopping amplitudes respectively. In what follows, we shall denote two filled sites across a rung as a {\it bosonic dipole}; the Fock state schematically shown in Fig.\ \ref{fig1}(a) is a state with maximal bosonic dipole number ($N_d^{\rm max} =N/4$ for ladders with $N$ sites)  at half-filling. The effect of the magnetic flux~($\Phi$) is incorporated by choosing a gauge so that a Peierls phase $\pm \phi/2$, with 
$\phi = 2 \pi \Phi/\Phi_0$ (where $\Phi_0= 2 \pi \hbar c/e, $ is the flux quantum, $c$ is the speed of light and $e$ denotes the effective boson change) appears in the intra-leg hopping term (Fig.\ \ref{fig1}(a)). The presence of a magnetic field breaks time-reversal symmetry which is crucial for the possible emergence of a chiral current. 

{\em Signatures of Meissner phase:-}
\label{chiral} 
The current operator along the legs and the rungs of the ladders is given by 
\begin{equation}
\begin{split}
    \hat J_{j, \sigma}^{\parallel} &= iJ\left[e^{-i\phi_\sigma/2}\hat{b}_{j,\sigma}^\dagger\hat{b}_{j+1,\sigma}-e^{i\phi_\sigma/2}\hat{b}_{j+1,\sigma}^\dagger\hat{b}_{j,\sigma}\right] \\
    \hat J_j^{\perp} &= iJ_\perp\left[\hat{b}_{j,A}^\dagger\hat{b}_{j,B}-\hat{b}_{j,B}^\dagger\hat{b}_{j,A}\right], 
\end{split}
\label{lcur1}
\end{equation}
respectively, where we choose $\phi_A = \phi$ and $\phi_B = -\phi$. We note that $J_{j, \sigma}^{\parallel}$ is a gauge-dependent quantity, but its expectation value between any state is gauge-independent~\cite{Moller_2010}. The chiral current, which shall be the focus of our study, can be obtained from Eq.\ \ref{lcur1} as $\hat J_c =\sum_j (\hat J_{j, A}^{\parallel}-\hat J_{j, B}^{\parallel})/L$, where $L$ is the length of the leg. 
A finite expectation value of $\hat J_c$ and vanishing expectation value $\hat J_j^{\perp}$ is a marker for the Meissner phase  \cite{Hugel_lad_2014, Atala2014, Greschner_lad_2015, Greschner_lad1_2015, Greschner_lad_2016}. 
We analyze $J_c(mT)=\langle \psi(mT)|\hat J_c|\psi(mT)\rangle$ for the driven ladder, where $|\psi(mT)\rangle$ denotes the wavefunction of the ladder after $m$ drive cycles. For this purpose, unless explicitly mentioned otherwise, we carry out exact diagonalization (ED) on finite ladders with $N=2L=20$ sites and use periodic boundary conditions (PBC) (see Ref.\ \cite{supplementary} for details). 
\begin{figure}[t!]
    \centering
    \includegraphics[width=1\columnwidth]{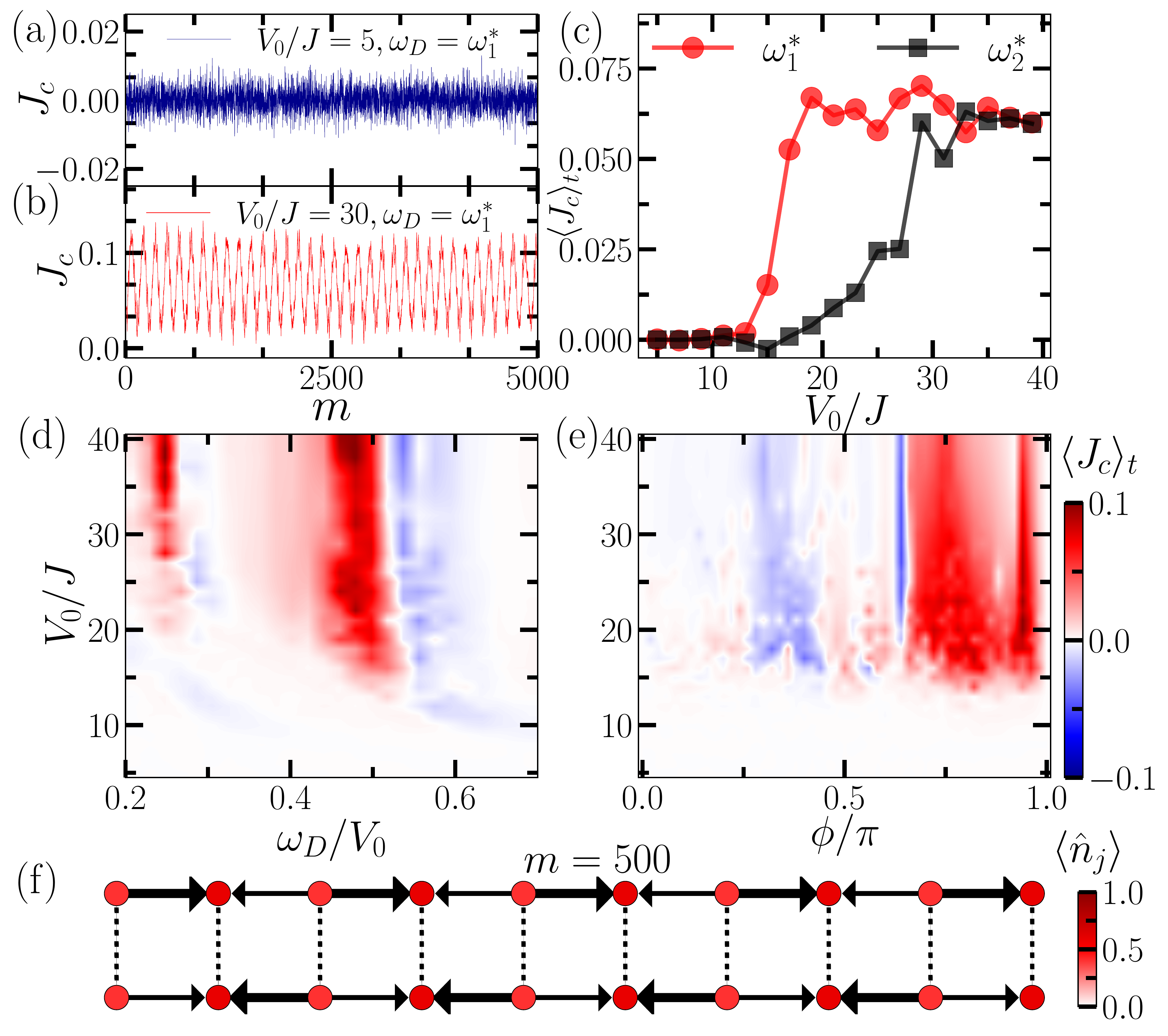}
    \caption{ Plot of $J_c(mT)$ as a function of $m$ for $V_0/J = 5$ (a) and $30$(b) respectively at $\omega_D = \omega_1^{\ast}= V_0/(2\hbar)$. (c) Plot of $\langle J_c \rangle_t$ as a function of $V_0/J$ for $\phi = 0.75\pi$ and $\omega_D=\omega_1^{\ast}$ (red circle) and $\omega_2^{\ast}$ (black sqaure). (d) Plot of $\langle J_c \rangle_t$ as a function of $\hbar \omega_D/V_0$ and $V_0/J$ for $\phi = 0.75\pi$. (e) Plot of $\langle J_c \rangle_t$ as a function of $\phi$ and $V_0/J$ for $\omega_D = \omega_1^{\ast}$. (f)  Plot of expectation value of the local current along each bond and the average particle density $\langle \hat{n}_i\rangle$ at each site for $V_0/J=30$, $\omega_D=\omega_1^{\ast}$ and $\phi = 0.75\pi$ at $m=500$. The width of the arrow represents the amplitude of the current while the dotted lines indicate bonds with zero current; the shown current pattern indicates a finite $J_c(mT)$. For all plots, $V_\parallel/J = 1$, and $J_\perp/J = 1$. See text for details.}
    \label{fig:chiral_curr}
\end{figure}
We choose an initial Fock state $|\psi(0)\rangle = \left| \begin{array}{cccccccccc}
1 & 0 & 1.~.~. & 1 & 0\\
1 & 0 & 1.~.~. & 1 & 0 
\end{array}\right\rangle$.  
The translationally symmetric partner of the above state is
$|\tilde \psi(0)\rangle = T_x |\psi(0)\rangle = \left| \begin{array}{cccccccccc}
 0 & 1.~.~. & 1 & 0 & 1\\
 0 & 1.~.~. & 1 & 0 & 1
\end{array}\right\rangle$,
where $T_x$ denotes the global translation operator which moves all particles on both legs by one site. $|\psi(0)\rangle$ and $|\tilde \psi(0)\rangle$ are half-filled states with maximum number of bosonic dipoles; they can be thought to be symmetric partners where the associated $Z_2$ symmetry corresponds to the occupation of odd and even sites of the legs of the ladder. 

To ascertain the presence of the Meissner phase, We first numerically compute $J_c(mT)$ using ED. A plot of $J_c(mT)$ as a function of $m$ is shown for $V_0/J = 5$ and $V_0/J = 30$ in Fig.~\ref{fig:chiral_curr}(a) and (b) respectively with $\omega_D = \omega_1^{\ast}=V_0/(2\hbar)$. For $V_0/J=5$, $J_c(mT)$ oscillates with a small amplitude about zero mean; in contrast, its oscillation is positively biased and occurs with a larger amplitude for $V_0/J=30$ at these drive frequencies. These oscillations have long time period and do not show any sign of decay at least for $m \sim 20000$ as checked numerically showing the presence of a long prethermal regime.

A plot of time-averaged value of $J_c(mT)$, given by $\langle J_c\rangle_t = \sum_{m=m_1}^{m_1+M} J_c(mT)/M$ with $m_1=500$ and $M=4500$, is shown in Fig.\ \ref{fig:chiral_curr}(c) as a function of $V_0/J$ for $\omega_D= \omega_1^{\ast}$ (red
circle) and $\omega_2^{\ast}=\omega_1^{\ast}/2$(black sqaure). It displays a sharp crossover from zero time-averaged Meissner current to a finite one around $V_0/J \simeq 15$ indicating the presence of a prethermal Meissner phase at higher drive amplitudes. To obtain a complete picture of the situation, we plot $\langle J_c \rangle_{t}$ as a function of $V_0/J$ and $\hbar \omega_D/V_0$ for a fixed $\phi = 0.75\pi$  in Fig.~\ref{fig:chiral_curr}(d). 
We find that $\langle J_c \rangle_{t}$ becomes large for $V_0/J \gg 1 $ and only at special frequencies $\omega_D=\omega_n^{\ast} = V_0/(2 n\hbar)$ where $n$ is an integer; it is small or close to zero for all other parameter regimes.  A plot of $\langle J_c \rangle_{t}$ as a function of $\phi$ and $V_0/J$ for fixed $\omega_D=\omega_1^{\ast}$, shown in Fig.~\ref{fig:chiral_curr}(e), indicates a direction reversal of the chiral current as a function of $\phi$  at fixed $V_0/J$. This is usually seen in flux ladders at equilibrium \cite{Greschner_lad1_2015}, but has not been observed in the driven ladder. To visualize this phenomenon, we show the $\langle\hat{n}_j \rangle= \langle \hat{b}_{j,\sigma}^\dagger \hat{b}_{j, \sigma}\rangle$ on ladder sites along with the local current direction in Fig.~\ref{fig:chiral_curr}(f) for $m=500$ and $\omega_D=\omega_1^{\ast}$. The pattern indicates vanishing of the current along the rung and its alternating pattern along the two legs indicating a nonzero $J_c(mT)$. These results constitute a prethermal Floquet realization of the Meissner phase in such driven flux ladders.

{\em FPT and numerical signature of HSF:}
\label{numerics} 
 To obtain a semi-analytic understanding of the origin of the Meissner current, we analyze the system using the Floquet perturbation theory~(FPT)~\cite{Sen_2021}. Within FPT, in the regime $V_0 \gg J,~J_\perp,~V_\parallel$, the leading order Hamiltonian is $\hat{H}_0(t)$. The corresponding time evolution operator is $U_0(t, 0) = \mathcal{T}\text{exp}\left[-i\int_0^t \hat{H}_0(t^\prime) dt^\prime/\hbar\right]$, where $\mathcal{T}$ is the time ordering operator. It can be shown that $U_0(T,0)=I$ where $I$ is the identity operator \cite{supplementary}; thus Floquet Hamiltonian $\hat{H}_F^{(0)}=0$ at this order. The first-order correction to the evolution operator is given by $U_1(T, 0) = (-i/\hbar)\int_0^T dt\left[U_0(t,0)^\dagger \hat{H_1} U_0(t,0)\right]$ and yields the Floquet Hamiltonian \cite{supplementary}
\begin{eqnarray}
\hat{H}_F^{(1)} &&= \frac{i \hbar}{T} U_1(T,0) = -J\sum_{i, \sigma = A, B} \left[(1- \hat{\chi}_{j \sigma}^2) + g(\gamma_0) \chi_{j \sigma}^2 \right] \nonumber \\
&& \times \left(~e^{-i\phi/2} \hat{b}_{j,A}^{\dagger}\hat{b}_{j+1,A}+e^{i\phi/2} \hat{b}_{j,B}^{\dagger}\hat{b}_{j+1,B}+{\rm h.c.}\right) \nonumber \\
&& -J_\perp\sum_{j, \sigma = A, B}\hat{b}_{j,\sigma}^{\dagger}\hat{b}_{j,\bar{\sigma}}+H.c.+ V_{\parallel}\sum_{j} \hat{n}_{j \sigma} \hat n_{j+1 \sigma},
\label{eq:ham_H_F_1}
\end{eqnarray}
where $g(\gamma_0) = \sin \gamma_0\exp(-i\gamma_0\hat{\chi}_{j \sigma})/\gamma_0$, $\hat{\chi}_{i \sigma} = \hat{n}_{i+1, \overline{\sigma}} - \hat{n}_{i, \overline{\sigma}}$, $\overline \sigma= B,A$ for $\sigma=A,B$, and $\gamma_0 = V_0T/4\hbar$. We note that hopping processes with $\chi_{j \sigma}=\pm 1$ {\it change the number of bosonic dipoles} (Fig.\ \ref{fig1}(b)) while those with $\chi_{j \sigma}=0$ conserve it (Fig.\ \ref{fig1}(c)). For $g(\gamma_0)=0$, {\it i.e.}, $V_0T/(4 \hbar)= n \pi$, where $n$ is an integer, only bosonic dipole conserving hopping processes are allowed at the first order \cite{flhsf1_1,flhsf1_2,supplementary}.  This leads to an approximate emergent bosonic dipole number conservation which controls the dynamics up to a large prethermal timescale in the large drive amplitude regime.  A detailed analysis of $\hat{H}_F^{(1)}$ at these special drive frequencies $\omega_D = \omega_n^{\ast}= V_0/(2 n \hbar)$ indicates that this additional conservation leads to strong HSF for $\hat{H}_F^{(1)}$ \cite{supplementary}. We note that the higher-order terms in the Floquet Hamiltonian $\hat{H}_F$ do not exhibit any fragmentation, leading to the eventual restoration of ergodicity. However, in the regime $V_0 \gg V_{\perp}, ~J,~ J_{\perp}$, the dynamics is governed by $\hat{H}_F^{(1)}$ up to an exponentially large prethermal timescale~ \cite{flhsf1_1,flhsf1_2, flhsf2_1,flhsf2_2,mori1_1,mori1_2} where the effect of HSF is prominent.  
\begin{figure}[t!]
    \centering
    \includegraphics[width=1\columnwidth]{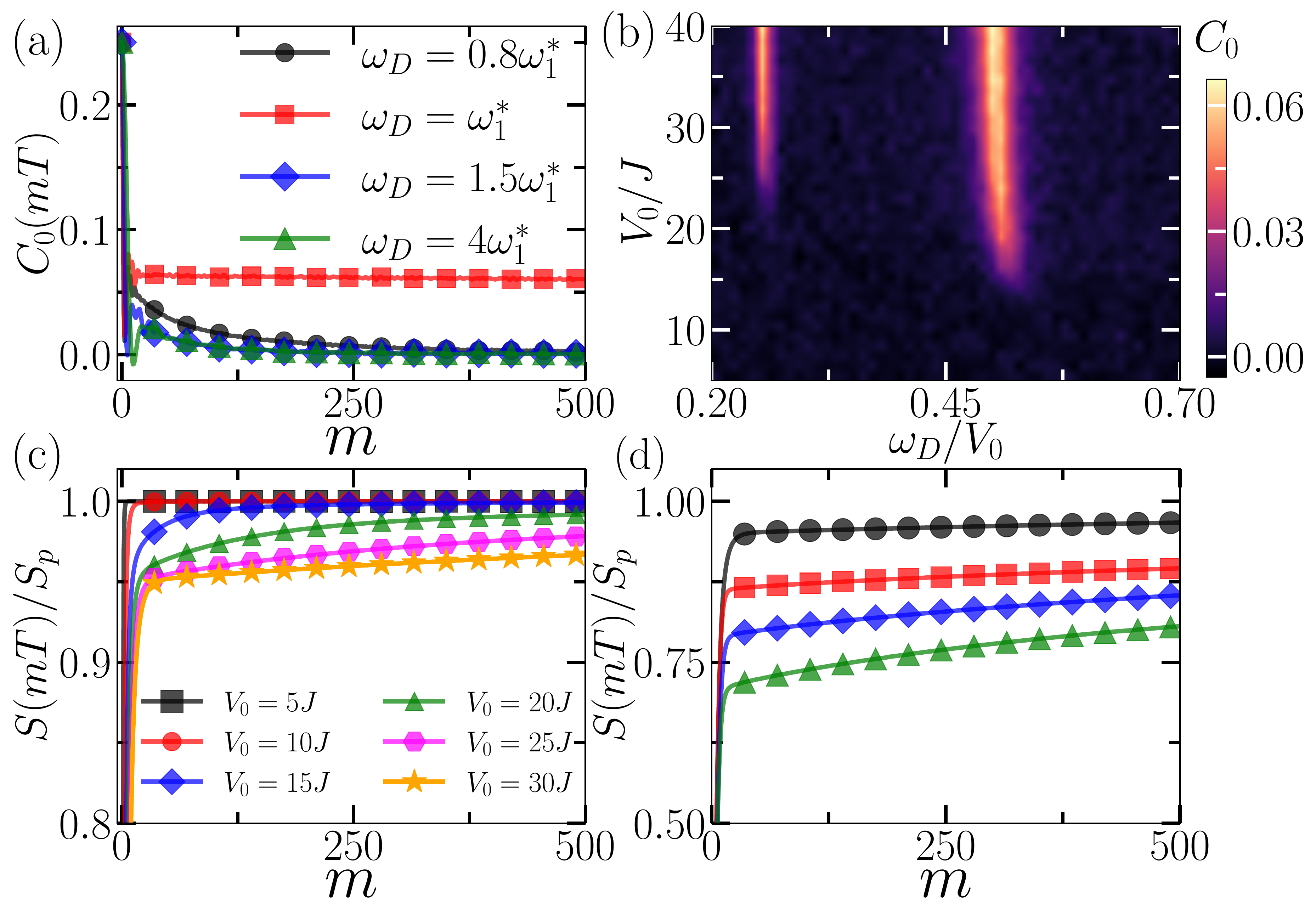}
\caption{ (a) Plot of $C_0(mT)$ as a function of $m$ for $V_0/J=30$  for different values of $\hbar \omega_D/J$. (b) Plot of $C_0(mT)$ for $m=2000$ as a function of $V_0/J$ and $\hbar \omega_D/V_0$ for $N=16$ sites. For these plots $C_0(mT)$ has been calculated using open boundary conditions and after averaging over (a) $60$ and (b) $20$ random initial states. (c) Plot of $S(mT)/S_p$ as a function of $m$ for several $V_0/J$ and $\omega_D=\omega_1^{\ast}$ for a random initial Fock state chosen from the largest fragment of $\hat{H}_F^{1)}$ with HSD $53760$. (d) Plot of $S(mT)/S_P$ is plotted as a function of $m$ for several random initial states. The corresponding fragment of $\hat{H}_F^{1)}$ has HSD $53760$ (black circles), $26880$ (red square), $23040$ (blue diamonds), and $6720$ (green triangles). For all plots, $V_0/J = 30$ and $\omega_D =\omega_1^{\ast}$. The plots (c) and (d) correspond to averaging  over $60$ random initial states from the same fragment. For all the calculations, we have chosen $V_\parallel/J=1$, $\phi=0.75\pi$, and $J_\perp/J=1$.}
    \label{fig:entropy_corr}
\end{figure}

To demonstrate the effect of such an emergent HSF on dynamics, we focus on density-density autocorrelation and half-chain entanglement of the driven ladder. A plot of the autocorrelation given by $C_{j \sigma}(mT) = [\frac{1}{2} \langle \varphi(0)|\left[\hat{n}_{j \sigma}(mT)-1/2\right] \left[\hat{n}_{j\sigma}(0)-1/2\right]|\varphi(0)\rangle +{\rm h.c.}]$
is shown in Fig.\ \ref{fig:entropy_corr}(a) and (b)  for $\sigma =A$ and $j=0$ (end site of the ladder). Here $|\varphi(0)\rangle$ is a random initial Fock state. For an ergodic driven system, $C_{j \sigma}(mT) \to 0$ for $m \to \infty$ in accordance with Floquet ETH; however, for a system exhibiting HSF, it is expected to remain above a lower bound known as the Majur bound \cite{MAZUR_1969, Jesko_2020, Sala_2020, Tibor_2020}. We find, from Fig.\ \ref{fig:entropy_corr}(a) and (b), $C_0 \equiv C_{0A}$ remains finite up to a large prethermal time for large $V_0/J$ and at $\omega_D= \omega_n^{\ast}$ for $n=1,2$. For all other parameter regimes, it vanishes for large $m$. This feature is consistent with existence of prethermal HSF at $\omega_D=\omega_n^{\ast}$ and large $V_0/J$. Importantly, a finite $J_c(mT)$ appears  in the same parameter regime at which prethermal HSF occurs (compare Fig.~\ref{fig:chiral_curr} (d) and Fig.~\ref{fig:entropy_corr}(b)).

 The lower panels of Fig.\ \ref{fig:entropy_corr} show the bipartite entanglement entropy after $m$ drive cycles: $S(mT)= -{\rm Tr}\left[\rho_A(mT) \ln (\rho_A(mT) )\right]$, where $\rho_A(mT) = {\rm Tr}_B(|\psi(mT)\rangle\langle\psi(mT)|)$ is the reduced density matrix for the subsystem $A$. Here we have chosen the subsystem $A$ to be half of the ladder constructed by a vertical cut parallel to its rungs. For Fig.\ \ref{fig:entropy_corr}(c), the initial state is chosen to be a Fock state from the largest fragment of $\hat{H}_F^{(1)}$, while for  Fig.\ \ref{fig:entropy_corr}(d) we have chosen several Fock states randomly from different fragments of $\hat{H}_F^{(1)}$ with varying Hilbert space dimension (HSD). For an ergodic system $S(mT)$ is expected to approach its Page value $S_{p} \sim \ln {\mathcal D}$, where ${\mathcal D}$ denotes the HSD; thus $S(mT)/S_{p} \to 1$ for large $m$. However, for fragmented system, $S(mT)$ approaches the Page value of the fragment to which the initial state belongs; this is denoted by $S_p^f \sim \ln {\mathcal D}_f$, where ${\mathcal D}_f$ is the HSD of the fragment. Thus, $S(m T)/S_p < 1 $ for systems with strong HSF \cite{Sala_2020}. In Fig.\ \ref{fig:entropy_corr}(c), we find that at $\omega_D=\omega_1^{\ast}$, $S(mT)/S_p <1$ for large $V_0/J$ while it approaches $1$ for small $V_0/J$. Moreover, in Fig.\ \ref{fig:entropy_corr}(d), we find that $S(mT)/S_p$
approaches different steady state values when the initial state is drawn from different fragments of $\hat{H}_F^{(1)}$ clearly showing breakdown of ergodicity at prethermal times for $\omega_D=\omega_1^{\ast}$ and $V_0/J \gg 1$.



{\it Emergent Meissner current:} To explain the relation between the prethermal HSF and the emergent Meissner phase, we note that the initial Fock states $|\psi(0)\rangle$ and $|\tilde \psi(0)\rangle$ are zero energy eigenstates (frozen states) of $\hat{H}_F^{(1)}$ having $N_d^{\rm max}$ dipoles. The dynamics of these states are therefore governed, up to a large prethermal timescale, by higher-order terms in $\hat{H}_F$ which are suppressed for large drive amplitudes. This restricts the spread of the system wavefunction in Hilbert space \cite{flhsf1_1}. 

\begin{figure}[t!]
    \centering
    \includegraphics[width=1\columnwidth]{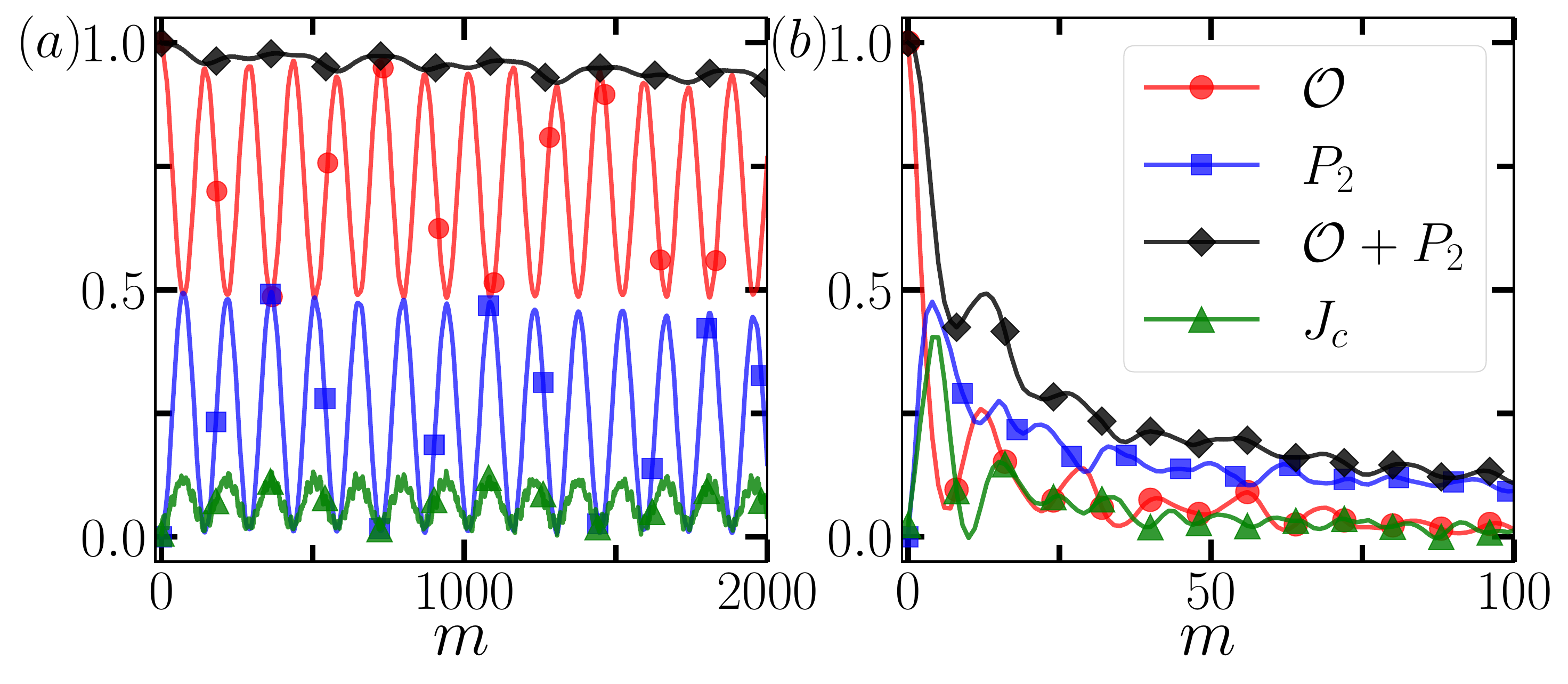}
    \caption{ Plot of $\mathcal{O}(m) = \mathcal{O}_1(m)+\mathcal{O}_2(m)$, $P_2(m)$, $\mathcal{O}(m)+P_2(m)$, and $J_c(mT)$ as a function of $m$ for (a) $\omega_D = \omega_1^*$ and (b) $\omega_D = 0.8\omega_1^*$, respectively. For all plots $J_\perp/J = 1,~V_0/J=30,~V_\parallel/J=1$, and $\phi=0.75\pi$.}
    \label{fig:overlap}
\end{figure}

To verify this assertion, we study the overlaps $\mathcal{O}_1(m) = |\langle \psi(mT)|\psi(0)\rangle|^2$, $\mathcal{O}_2(m) = |\langle \psi(mT)|\tilde\psi(0)\rangle|^2$
and $P_2(m) = \sum_i |\langle \psi(mT)|n_i\rangle|^2$, where $|n_i\rangle$ denotes a Fock state with $N_d^{\rm max}-2$ bosonic dipoles (see Ref.\ \cite{supplementary} for details). A plot of these overlaps shown in Fig.\ \ref{fig:overlap}(a) clearly indicates that $|\psi(mT)\rangle$, for $m\le 2000$, has almost unit total overlap with the states in these sectors; thus, it is almost entirely localized within these sectors till a large prethermal timescale. This indicates that such a wavefunction can be written as $|\psi(mT)\rangle \simeq a(mT) |\psi(0)\rangle + b(mT) |\tilde \psi(0)\rangle +\sum_{i} c_i(mT) |n_i\rangle$ leading to a finite expectation value of the chiral current since $\langle n_i|\hat J_c |n_i\rangle$, $\langle \psi(0)|\hat J_c |n_i\rangle$, and $\langle \tilde \psi(0)|\hat J_c |n_i\rangle \ne 0$. We note that such a localization does not occur in the absence of prethermal HSF as can be seen from  Fig.\ \ref{fig:overlap}(b); in this case $|\psi(mT)\rangle$ spreads throughout the entire Hilbert space leading to rapid decay of $\mathcal{O}_{1,2}(m)$ and $P_2(m)$ within $m\le 100$ cycles. This feature is consistent with Floquet ETH and leads to $\langle J_c\rangle_t=0$.

\begin{figure}[t!]
    \centering
    \includegraphics[width=1\columnwidth]{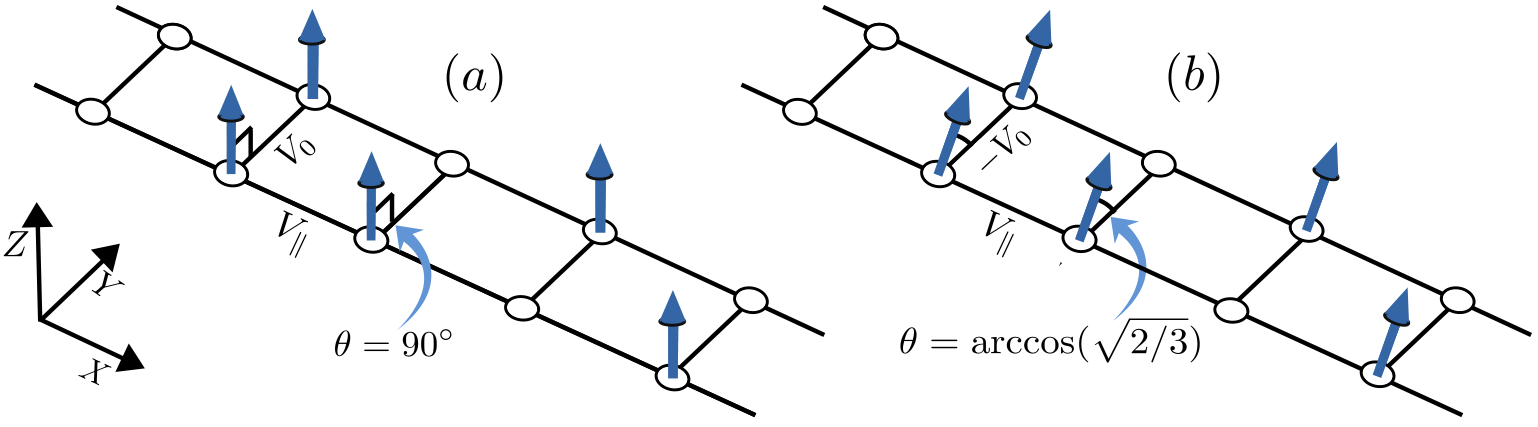}
    \caption{ Schematic representation of the proposed experimental setup showing the alignment of atomic dipole moments when the interaction along the rung is (a) $V_0$ and (b) $-V_0$ respectively, while keeping $V_\parallel$ fixed.}
    \label{fig:toy_plot_exp}
\end{figure}

{\it Experiments:} The proposed model can be realized by considering dipolar atoms in an optical lattice having a two-leg ladder geometry and in the limit of large lattice depth \cite{Lahaye_2009, Baranov2012, Chomaz_2023}.  Experimentally, to realize fluxes through plaquettes, it is convenient to assign phases to $J_{\perp}$ through laser-assisted tunneling ~\cite{Jaksch_2003, Gerbier_2010, Jim_2012, Struck_2012, Ketterle_2013, Monika_2013, Atala2014, Aidelsburger2015, Tai2017, impertro2025}; this amounts to a different gauge choice but is physically equivalent to Eq.\ \ref{eq:Ham}~\cite{Rey_2022}. The dipolar interaction between the atoms is given by $V_{d} = \frac{V_0 a^3}{r^3} (1-3\text{cos}^2\theta)$, where $a$ is the lattice spacing, $r$ is the distance between the dipoles, $\theta$ is the angle between the dipole moment and the line joining the dipoles, and $V_0$ is the interaction between the neighboring dipoles with $r=a$ and $\theta=90^{\circ}$. This interaction can be tuned by varying $\theta$ using an external magnetic field in the following manner.  Initially, all dipoles are aligned along the $Z$-direction (Fig.~\ref{fig:toy_plot_exp}(a)), and the distances between dipoles along the legs and rungs are chosen so that $V_0/V_\parallel\simeq30$. In this configuration, both leg and rung interactions are repulsive. We then apply an external magnetic field to rotate the dipoles within the $YZ$-plane, ensuring that the angle between dipoles along the legs remains $90^{\circ}$, while the angle between dipoles across the rungs is set to  $\theta = \arccos(\sqrt{2/3})$ (Fig.~\ref{fig:toy_plot_exp}(b)). This leads to attractive rung interaction: $V_\perp=-V_0$. The diagonal interaction within a plaquette always remains smaller than  $V_\parallel$. Thus, a periodic variation of the external magnetic field with frequency $\omega_D$, leads to a periodic variation of $V_\perp$ between $V_0$ to $-V_0$ keeping $V_{\parallel}$ and $J$ fixed. We propose measurement of Meissner current in this setting \cite{Atala2014, impertro2025} and predict that it will be finite at the special drive frequencies $\hbar \omega_D/V_0=1/(2n)$ and large drive amplitude regime; for other parameter regimes, $\langle \hat J_c \rangle_t=0$. 
 
{\em Conclusions.-} In conclusion, we have studied driven flux ladders and have shown the existence of a prethermal Meissner phase in such systems. Our analysis ties the presence of this phase to the emergent prethermal HSF in such ladders at large drive amplitude and special drive frequencies providing an yet unexplored connection to these two well-known phenomena. We have also discussed experimental setup using dipolar atoms in optical lattices that can potentially realize this phenomenon.   

{\em Acknowledgement.-}
T.M. acknowledges support from the Science and Engineering Research Board (SERB), Govt. of India, through project No. MTR/2022/000382 and STR/2022/000023.
KS thanks DST for support through SERB project JCB/2021/000030.

\bibliography{ref}
\clearpage
\onecolumngrid
\begin{center}
\textbf{Supplementary materials for 
 ``Floquet realization of prethermal Meissner current on two-leg flux ladder "}
\end{center}

    In this supplementary material, we provide a detailed description of the numerical technique used for time evolution and the derivation of the first-order Floquet Hamiltonian using Floquet perturbation theory (FPT). We also numerically analyze the structure of Hilbert space fragmentation~(HSF) and analytically determine the number of frozen states of the first-order Floquet Hamiltonian. Finally, we provide a more detailed analysis of the dynamics of the driven ladder starting from a frozen state and discuss the property of the chiral current in light of such dynamics.

\vspace{5mm}

\twocolumngrid
\section{Numerical Method for Time Evolution}
\label{numerics}
In this section, we discuss in detail the time evolution procedure for the square pulse protocol. The stroboscopic evolution operator is given by
\begin{equation}
    U(T, 0) = \text{exp}(-i\hat{H}_+T/2)\text{exp}(-i\hat{H}_-T/2)
\end{equation}
where, $\hat{H}_\pm = \hat{H}(V_\perp = \pm V_0)$. We note that for any even or odd integer $m$, the time evolution of the quantum state can be written in terms of $H_{\pm}$ as 
\begin{equation}
    |\psi((m+1)T/2)\rangle = \text{exp}(-i\hat{H}_\alpha T/2)|\psi(mT/2)\rangle
\end{equation}
where $\alpha=-,~\text{or}~+$ for even or odd $n$. 

We use the Krylov subspace method~\cite{krylov} for the time evolution of the quantum state. The Krylov subspace $\mathcal{K}$, for $H= H_{\alpha}$ is spanned by the vectors 
\begin{eqnarray} 
\hat{Q}_n^{\alpha}=(|q_0^{\alpha}\rangle,~ |q_1^{\alpha}\rangle\,~.~.~.~.~|q_n^{\alpha} \rangle)^T.
\end{eqnarray}
The overlap of these basis states with the wavefunction $|\psi(mT/2)\rangle$ is given by 
\begin{eqnarray} 
c_i^{\alpha} &=& \langle \psi(mT/2)|q_i^{\alpha} \rangle = \langle \psi(mT/2)|(\hat{H}_\alpha)^i|\psi(mT/2)\rangle.
\end{eqnarray} 
Thus, the quantum state at time $(m+1)T/2$ can be well approximated in the $n+1$ dimensional Krylov space. The projection operator which projects the original quantum state to Krylov subspace is defined as
\begin{equation}
    \hat{P_n^{\alpha}} = \sum_{i=0}^n|q_i\rangle\langle q_i| = \hat{Q}_n^{\alpha \dagger} \hat{Q}_n^{\alpha} 
\end{equation}
In terms of these operators, the time-evolved quantum state can be written as
\begin{eqnarray}
|\psi((m+1)T/2)\rangle &=& e^{-i\hat{H}_\alpha T/2}|\psi(mT/2)\rangle \nonumber \\ 
&=& \hat{Q}_n^{\alpha\dagger} \hat{Q}_n^{\alpha}e^{-i\hat{H}_\alpha T/2}\hat{Q}_n^{\alpha \dagger} \hat{Q}_n^{\alpha}|\psi(mT/2)\rangle \nonumber \\
    &=& \hat{Q}_n^{\alpha\dagger} e^{-i\mathcal{\hat{H}}_\alpha T/2}\hat{Q}^\alpha_n|\psi(mT/2)\rangle,
\end{eqnarray}
where $\mathcal{\hat{H}}_\alpha = \hat{Q}_n^{\alpha} \hat{H}_\alpha \hat{Q}_n^{\alpha \dagger}$ and $\alpha=\pm$ depending on whether $m$ is even or odd. Since the Hamiltonian $\hat{H}_\alpha$ is a hermitian matrix,  $\mathcal{\hat{H}}_\alpha$ can be represented by a $(n+1) \times (n+1)$ tridiagonal matrix. Since $n$ is significantly smaller than the Hilbert space dimension of $\hat{H}_\alpha$, one can easily compute $\text{exp}(-i\mathcal{H}_\alpha T/2)$ using the standard procedure. 

A convergent result for the wavefunction can be obtained by gradually increasing the dimension of the Krylov subspace and for better convergence $T$ should be kept small. This method allows us to avoid the full diagonalization of the matrix and thus reduces the computational cost significantly. Therefore, we can access larger system sizes with the same computational resources. We have checked for smaller systems sizes that the result obtained from this method converges to that obtained from conventional ED retaining all states in the Hilbert space.

\section{Floquet Perturbation Theory(FPT)}
\label{appa1} 

In this section, we use FPT to compute the Floquet Hamiltonian for $V_0 \gg J, J_{\perp}, V_1$ using the square-pulse drive protocol given in the main text. In this limit, the zeroth order evolution operator (which becomes exact for $\hat{H}_1=0$) is given by,
\begin{equation}
    U_0(t, 0) = \mathcal{T}\text{exp}\left[-i\int_0^t \hat{H}_0(t^\prime) dt^\prime\right],
    \label{eq:flq1}
\end{equation}
where, $\mathcal{T}$ is the time ordering operator. For the square-pulse protocol given in the main text, $U_0(t, 0)$ takes the form
\begin{equation}
    \begin{split}
        U_0(t, 0) &= \text{exp}\left[-i\frac{V_0t\hat{O}}{\hbar}\right], ~~ \text{for} ~t\leq T/2 \\
        &=\text{exp}\left[-i\frac{V_0(T-t)\hat{O}}{\hbar}\right], ~~ \text{for} ~ t>T/2,
    \end{split}
\end{equation}
where, $\hat{O} = \sum \hat{n}_{i, A}\hat{n}_{i, B}$. We note that $U_0(T, 0) = I$, where $I$ denotes the identity operator, indicating that $\hat{H}_F^{(0)} = 0$. Thus the contribution to the Floquet
Hamiltonian ($\hat{H}_F$) comes from higher order terms in perturbation theory, which we now compute. 

To this end, we note that the first-order correction to $U_0$ is given by,
\begin{equation}
    U_1(T, 0) = -\frac{i}{\hbar}\int_0^T dt\left[U_0(t,0)^\dagger \hat{H_1} U_0(t,0)\right].
\end{equation}
The perturbative Hamiltonian contains two types of terms, namely the inter- and intra-leg hopping terms, and the intra-leg interaction term. Since the intra-leg interaction term in $H_1$ commutes with $U_0$, its 
contribution to $\hat{H}_F^{(1)}$ is trivially computed and leads to 
\begin{eqnarray} 
U_{1a}(T,0) &=& \frac{-i V_{\parallel}T}{\hbar} \sum_{j, \sigma = A, B} \hat{n}_{j, \sigma}\hat{n}_{j+1, \sigma} \nonumber\\
\hat{H}_F^{(1a)} &=& i\hbar U_{1a}/T = V_{\parallel} \sum_{j, \sigma = A, B} \hat{n}_{j, \sigma}\hat{n}_{j+1, \sigma}. \label{fl1}
\end{eqnarray}

In contrast, the kinetic term in $\hat{H}_1$ does not commute with $U_0$; these terms yield the non-trivial part of $\hat{H}_F^{(1)}$. To obtain these, we first consider the action of $U_0(t)$ on a 
general Fock state $|\alpha\rangle$. We find 
\begin{equation}
    \begin{split}
        U_0(t, 0)|\alpha\rangle &= \text{exp}\left[-i\frac{V_0tP_\alpha}{\hbar}\right]|\alpha\rangle,~~\text{for} ~ t\leq T/2 \\
        &=\text{exp}\left[-i\frac{V_0(T-t)P_\alpha}{\hbar}\right]|\alpha\rangle,~~\text{for} ~ t> T/2,
    \end{split}
    \label{eq:U0phase}
\end{equation}
where, $\sum \hat{n}_{i, A}\hat{n}_{i, B}|\alpha\rangle = P_\alpha|\alpha\rangle$. 

\begin{figure}[t!]
    \centering
    \includegraphics[width=1\columnwidth]{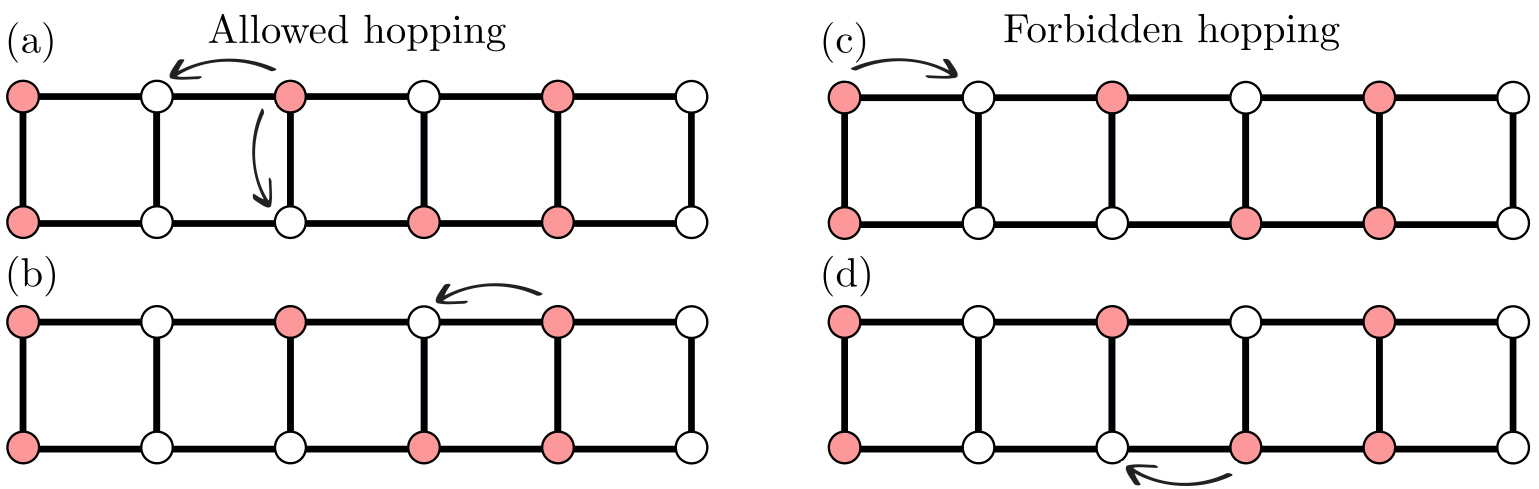}
    \caption{ Schematic representation of (a) inter-leg and (b) intra-leg hopping processes which connect states with same $P_{\alpha}$. (c) and (d) Intra-leg hopping processes which change $P_{\alpha}$; we note that there is no inter-leg hopping which connects states with different $P_{\alpha}$. The former processes ((a) and (b)) are allowed while the latter ones ((c) and (d)) are forbidden by $\hat{H}_F^{(1)}$ at special drive frequencies. See text for details. }
    \label{fig:toy_hop}
\end{figure}

We first consider the inter-leg hopping term in $H_1$. The action of this term results in hopping of a boson from leg $\sigma=A,B$ to another leg $\bar{\sigma}= B,A$ while keeping the rung index $j$ unchanged as schematically shown in Fig.\ \ref{fig:toy_hop} (a). We consider the action of such inter-leg hopping term on the Fock state $|\alpha\rangle$: $ \hat{b}_{j, \sigma}^\dagger \hat{b}_{j, \bar{\sigma}}|\alpha\rangle = |\beta\rangle$. For such a process, clearly $P_\alpha=P_\beta$ (Fig.\ \ref{fig:toy_hop} (a)). Thus the contribution to the first-order Floquet Hamiltonian from this term can be easily computed and is given by 
\begin{eqnarray}
 \hat{H}_F^{(1b)} &=& -J_\perp\sum_{j, \sigma = A, B}(\hat{b}_{j,\sigma}^{\dagger}\hat{b}_{j,\bar{\sigma}}+ {\rm h.c.}).   \label{fl2}
\end{eqnarray}

In contrast, for the intra-leg hopping terms, the bosons hop  between neighboring sites of the same leg and $\sigma$ remains unchanged. Such a hopping process connects two Fock states $|\alpha\rangle$ and $|\beta\rangle$ such that $\hat{b}_{j, \sigma}^\dagger \hat{b}_{j+1, \sigma}|\alpha\rangle=|\beta\rangle$ as schematically shown in Fig.\ \ref{fig:toy_hop}(a-d). It can be easily seen that such intra-leg hopping processes connect Fock states $|\alpha\rangle$ and $|\beta\rangle$ such that 
\begin{eqnarray}
P_{\alpha} &=& P_{\beta}, \quad \text{if}~~\hat{\chi}_{j \sigma}|\alpha\rangle = 0 \nonumber \\
&=& P_{\beta} \pm 1,\quad \text{if} \quad \hat{\chi}_{j \sigma}|\alpha\rangle = \pm|\alpha\rangle \nonumber \\
\hat{\chi}_{j \sigma} &=& \hat{n}_{j+1, \bar{\sigma}} - \hat{n}_{j, \bar{\sigma}}.
\end{eqnarray}
Using this and the standard method provided in Ref.\ \onlinecite{rev11,flhsf1_1}, we find that the contribution to $U_1$ from the intra-leg hopping processes are given by,
\begin{eqnarray}
U_{1c} (T, 0) &=& U_{1c}^{(0)}(T,0) + U_{1c}^{(1)}(T,0) + U_{1c}^{(-1)}(T,0)\nonumber\\
U_{1c}^{(0)}(T,0) &=& \frac{iT J}{\hbar} \sum_{j, \sigma = A, B} \left(~e^{-i\phi/2} \hat{b}_{j,A}^{\dagger}\hat{b}_{j+1,A} \right. \nonumber\\
&&\left. +e^{i\phi/2} \hat{b}_{j,B}^{\dagger}\hat{b}_{j+1,B}+{\rm h.c.}\right) \nonumber\\
U_{1c}^{(\pm 1)}(T,0) &=& \frac{iT J}{\hbar} \sum_{i, \sigma = A, B}\frac{\sin \gamma_0}{\gamma_0} e^{\pm i\gamma_0}  \\
&& \times \left(~e^{-i\phi/2} \hat{b}_{j,A}^{\dagger}\hat{b}_{j+1,A}+e^{i\phi/2} \hat{b}_{j,B}^{\dagger}\hat{b}_{j+1,B}+{\rm h.c.}\right). \nonumber
\end{eqnarray} 
We note that $U_{1c}^{(0)}(T,0)$ represents hopping processes for which $P_{\alpha}= P_{\beta}$, {\it i.e.}, they correspond to $\hat \chi_{j \sigma}=0$. In contrast,  $U_{1c}^{(\pm 1)}(T,0)$
corresponds to hopping processes for which $\hat \chi_{j \sigma}=\pm1$. This allows one to combine these terms and write 
\begin{eqnarray} 
U_{1c}(T,0) &=&\frac{iT J}{\hbar} \sum_{i, \sigma = A, B}\frac{\sin{\gamma_0\hat{\chi}_{j \sigma}}}{\gamma_0\hat{\chi}_{j \sigma}} e^{i\gamma_0\hat{\chi}_{j \sigma}}  \\
&& \times \left(~e^{-i\phi/2} \hat{b}_{j,A}^{\dagger}\hat{b}_{j+1,A}+e^{i\phi/2} \hat{b}_{j,B}^{\dagger}\hat{b}_{j+1,B}+{\rm h.c.}\right) \nonumber\\
\hat{H}_F^{(1c)} &=& \frac{i \hbar}{T} U_{1c}(T,0) = -J \sum_{i, \sigma = A, B}\frac{\sin {\gamma_0\hat{\chi}_{j \sigma}}}{\gamma_0\hat{\chi}_{j \sigma}} e^{i\gamma_0\hat{\chi}_{j \sigma}} \label{fl3} \\
&& \times \left(~e^{-i\phi/2} \hat{b}_{j,A}^{\dagger}\hat{b}_{j+1,A}+e^{i\phi/2} \hat{b}_{j,B}^{\dagger}\hat{b}_{j+1,B}+{\rm h.c.}\right) \nonumber
\end{eqnarray}
where $\gamma_0 = V_0 T/(4 \hbar)$. 

Combining Eqs.\ \ref{fl1}, \ref{fl2} and \ref{fl3}, finally we find the first-order Floquet Hamiltonian as
\begin{eqnarray}
\hat{H}_F^{(1)} &=& i\hbar U_1(T, 0)/T=\hat{H}_F^{(1a)} + \hat{H}_F^{(1b)} + \hat{H}_F^{(1c)}  \label{eq:first_od_ham} \\
&=& -J\sum_{i, \sigma = A, B} \left[(1- \hat{\chi}_{j \sigma}^2) + g(\gamma_0) \chi_{j \sigma}^2 \right] \nonumber \\
&& \times \left(~e^{-i\phi/2} \hat{b}_{j,A}^{\dagger}\hat{b}_{j+1,A}+e^{i\phi/2} \hat{b}_{j,B}^{\dagger}\hat{b}_{j+1,B}+{\rm h.c.}\right) \nonumber \\
&& -J_\perp\sum_{j, \sigma = A, B}\hat{b}_{j,\sigma}^{\dagger}\hat{b}_{j,\bar{\sigma}}+H.c.+ V_{\parallel}\sum_{j} \hat{n}_{j \sigma} \hat n_{j+1 \sigma}, \nonumber \\
g(\gamma_0) &=& \frac{\sin{\gamma_0}}{\gamma_0}\exp(-i\gamma_0\hat{\chi}_{j \sigma}),\nonumber
\end{eqnarray}
where we have used the fact that  $\hat \chi_{j \sigma}$ can take values $0, \pm 1$. This leads to Eq. 3 of the main text.

From $\hat{H}_F^{(1)}$, it is easy to see that for $g(\gamma_0) = 0$ only hopping processes which corresponds to $\chi_{j \sigma}= 0$ are allowed. Since such processes conserves the number of bosonic dipoles (number of rungs with two bosons at their ends); $\hat{H}_F^{(1)}$ hosts an emergent bosonic dipole number ($N_d= \sum_j \hat{n}_{j,A}\hat{n}_{j,B}$) conservation. Such a conservation therefore occurs when $\sin \gamma_0/\gamma_0=0$, {\it i.e.} when $\gamma_0= n \pi$ or $\omega_D = \omega_n^* = V_0/2n\hbar$ with integer $n$. Such a conservation is approximate; we have numerically checked that exact $\hat{H}_F$ do not have such a conservation. In the next section, we show that this emergent conservation leads to Hilbert space fragmentation (HSF) for $\hat{H}_F^{(1)}$ at special drive frequencies $\omega_D=\omega_n^{\ast}$. 

 \begin{figure}[t!]
    \centering
    \includegraphics[width=1\columnwidth]{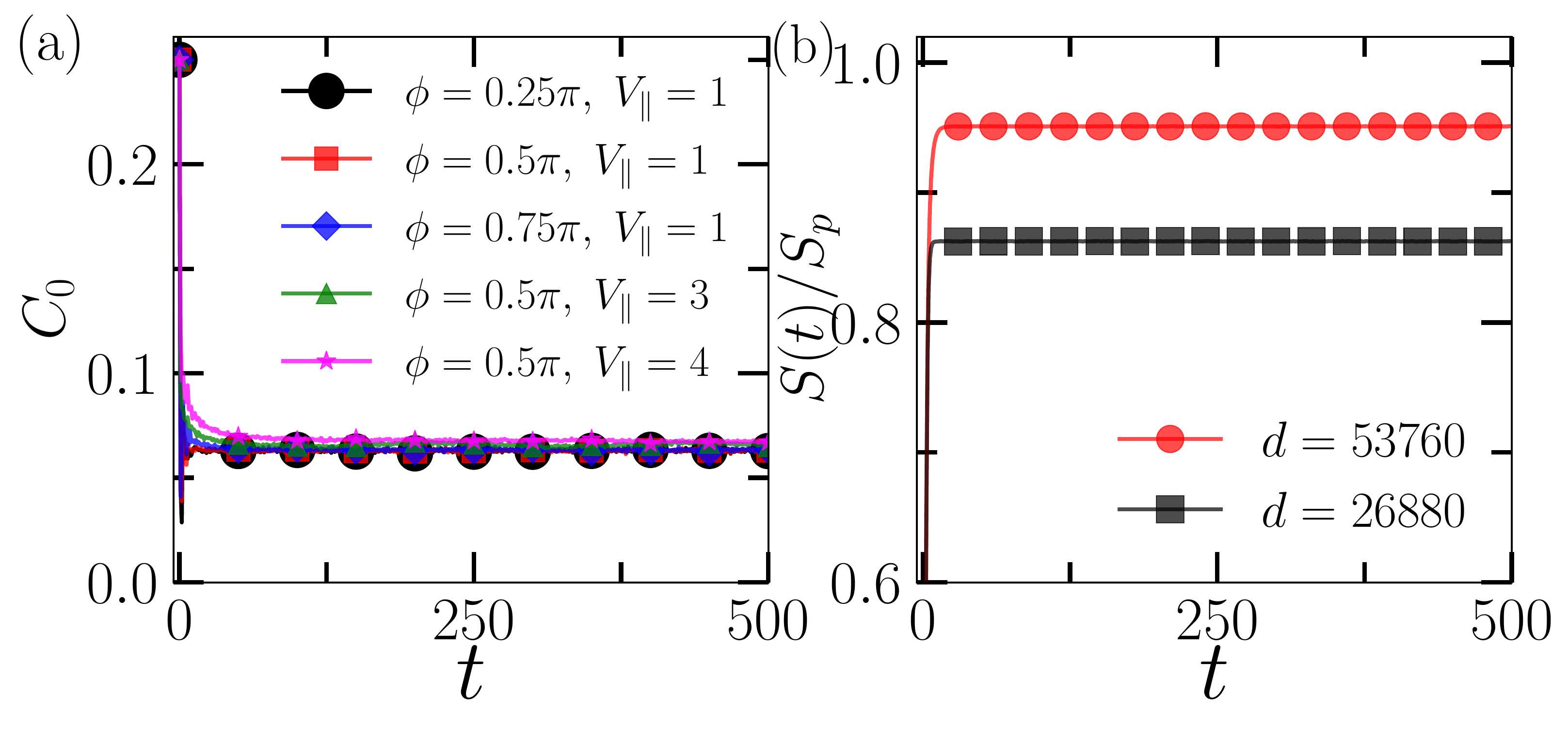}
    \caption{ (a) Plot of $C_0$ as a function of time (in units of $\hbar/J$) for different values of $\phi$ and $V_\parallel$. $C_0$ is computed by averaging over 30 random initial states, with open boundary conditions. (b) Plot of $S(t)/S_P$ as a function of time (in units of $\hbar/J$), calculated by taking an average over $30$ initial states chosen from different fragmented sectors. We consider a periodic boundary condition in this case and set $\phi = 0.5\pi$. Here, the dynamics are governed by the first-order Floquet Hamiltonian~($\hat{H}_F^{(1)}$). These calculations are done by considering half-filling with system size $L=10$, and the other parameters fixed to $V_\parallel/J=1$, and $J_\perp/J=1$.}
    \label{fig:first_od_ham_plot}
\end{figure}

\section{Emergent HSF for first-order Floquet Hamiltonian}
\label{appa2}

Here we analyze the properties of the first-order Floquet Hamiltonian~($\hat{H}_F^{(1)}$) at special drive frequencies for which $g(\gamma_0)=0$. First, we numerically calculate the dimension of the fragmented regions taking into account the constraints that appear in $\hat{H}_F^{(1)}$. We consider a system with $L=10$ rungs (i.e., $20$ lattice sites) and we study the system at half-filling. In this scenario, the total dimension of the Hilbert space is $184756$. Due to the constraint appearing in $\hat{H}_F^{(1)}$ the Hilbert space takes a block diagonal form with a largest fragmented Hilbert space dimension (HSD) of $53760$ followed by dimensions of $26880,~23040,~20160,~6720,~1440$ etc., along with several frozen states each of which constitutes an one dimensional fragment. Below, we demonstrates that appearence of such fragments constitutes a realization of strong HSF for $\hat{H}_F^{(1)}$ 
at special drive frequencies where $g(\gamma_0)=0$.

{\it Autocorrelation and Entanglement}: We study the time evolution of the system when the dynamics is governed by $\hat{H}_F^{(1)}$. The quantum state at a time $t$ is given by
\begin{equation}
    |\psi(t)\rangle = e^{-i\hat{H}_F^{(1)}t}|\psi(0)\rangle,
\end{equation}
where $|\psi(0)\rangle$ is the initial state at time $t=0$. In Fig.~\ref{fig:first_od_ham_plot}(a) we plot the autocorrelation~($C_0$) (defined in the main text) as a function of time starting from the random initial states, for different values of $\phi$ and $V_\parallel$. We consider open boundary conditions for our calculation and take the average over $30$ random initial states for each parameter values. We observe that $C_0$  saturates to a finite value. This indicates breakdown of eigenstate thermalization hypothesis (ETH) which, in the present case, serves as signature of HSF.  We note here that $C_0$ saturates to the same value for different $\phi$ and $V_\parallel$ indicating that the magnetic flux and the intra-leg interaction do not affect HSF. 

To further confirm the presence of fragmentation, we plot the bipartite entanglement entropy as a function of time as shown in Fig.~\ref{fig:first_od_ham_plot}(b). We choose the initial Fock states from two different fragmented sectors with Hilbert space dimensions $d=53760$ and $d = 26880$. For our calculations, we consider the periodic boundary condition, and we take an average over $30$ initial states randomly chosen from the same fragmented sector. In both cases, the bipartite entanglement entropy $S$ saturates below the page value~($S_p \sim 6.34$ for $L=20$ at half-filling). Moreover, for each fragmented sector, the entanglement entropy saturates to $S \sim \ln(D_f)$, where $D_f$ is the dimension of the fragmented sectors.  This is consistent with the presence of HSF for $\hat{H}_F^{(1)}$. 

\begin{figure}[t!]
    \centering
    \includegraphics[width=0.9\columnwidth]{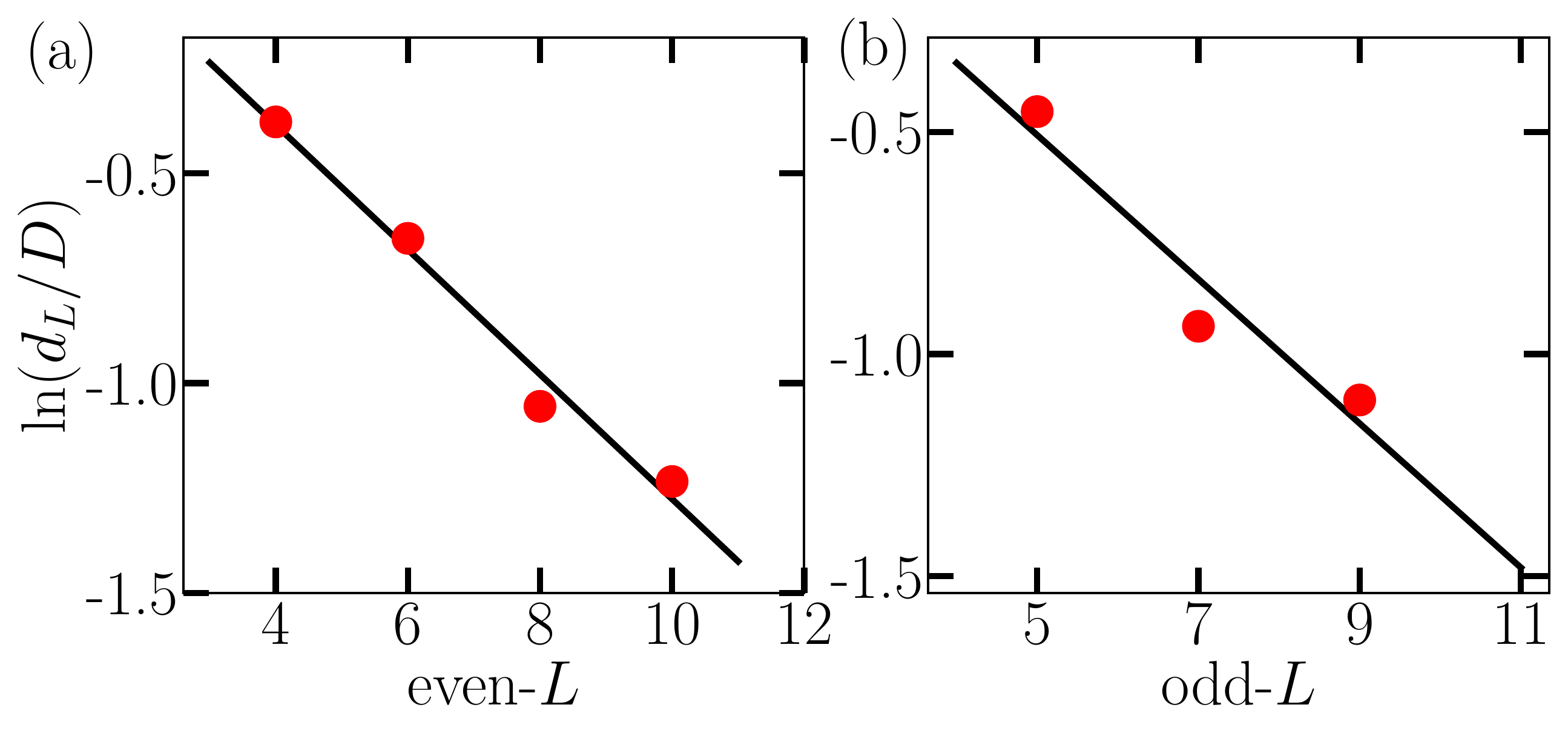}
    \caption{Plot of $\ln (d_L/{\mathcal D})$ at half-filling as a function of $L$ for even (a) and odd (b) $L$.}
    \label{fig:large_H_dim_vs_L}
\end{figure} 

\begin{figure*}[t!]
    \centering
    \includegraphics[width=1.6\columnwidth]{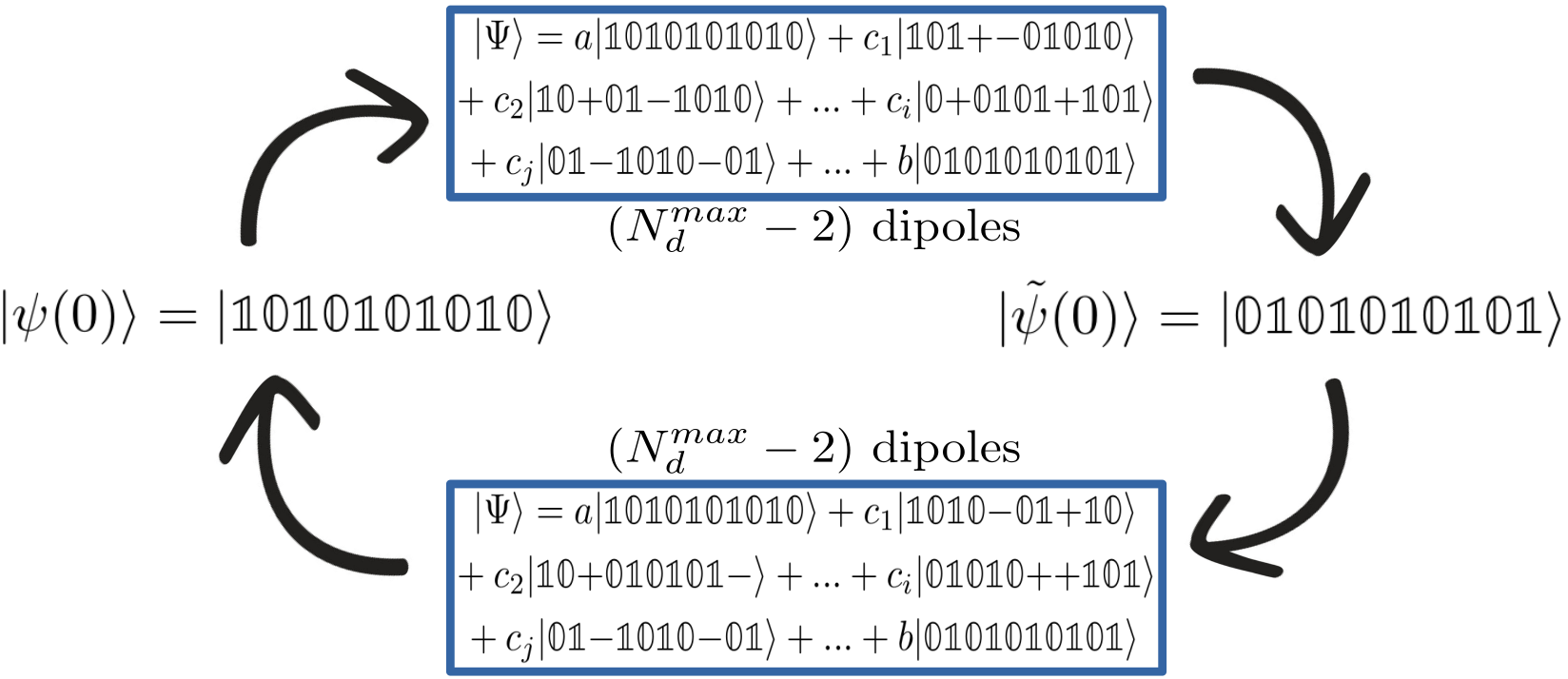}
    \caption{ Schematic representation of the dynamics representing coherent oscillations between two states $|\psi(0)\rangle
    $ and $|\tilde \psi(0)\rangle$ up to a long prethermal scale, mediated through ($N_d^{{\rm max}}-2$) dipole sector. The figure represents dynamics starting from $|\psi(0)\rangle$ with arrows above indicating the first half cycle and arrows below indicating the second half cycle.}
    \label{fig:toy_dynamics}
\end{figure*}

{\it Mapping to one-dimensional (1D) chain}: The structure of HSF for $\hat{H}_F^{(1)}$ can be conveniently studied through a mapping which maps Fock states of the ladder to that of a 1D chain.  For this mapping, it is convenient denote a state $|0 0\rangle$ which represents empty sites across a rung of the ladder. Similarly one defines states $|1 0\rangle$, $|01\rangle$ and $|11\rangle$ corresponding rungs with occupied $A$ leg, occupied $B$ leg, and double occupation respectively. Using this notation, one can map $|00\rangle \rightarrow |\mathbb{0} \rangle$, $|11\rangle \rightarrow |\mathbb{1} \rangle$, $|10\rangle \rightarrow |+ \rangle$ and $|01\rangle \rightarrow |- \rangle$. In this notation a bosonic dipole is represented by $|\mathbb{1}\rangle$.
This allows us to map an arbitrary Fock state of ladder geometry 
$\left( \left| \begin{array}{cccc}
1 & 0 & 1 & 0\\
1 & 0 & 0 & 1
\end{array}\right\rangle \right)$ to those for a 1D ($|\mathbb{1}\mathbb{0}+-\rangle$) without loss of generality.

{\it Frozen states}: The above map allows us to easily quantity quantify the number of frozen states, $N_{\rm frozen}$, of $\hat{H}_F^{(1)}$ at the special frequencies. We note that for the 1D chain Fock states $|\eta\rangle$ that only consist of $|\mathbb{1}\rangle$ and $|\mathbb{0}\rangle$ satisfy $\hat{H}_F^{(1)}|\eta\rangle= \epsilon_n |\eta\rangle$ where $\epsilon_n = V_{\parallel} n_s$ and $\sum_{j \sigma} \hat \hat n_{j \sigma} \hat n_{j+1 \sigma} |\eta\rangle= n_s|\eta \rangle$. Such states are not connected with the other Fock states through hopping processes at special drive frequencies; they constitute frozen states of $\hat{H}_F^{(1)}$. These are zero energy eigenstates states if $V_{\parallel}=0$ or $n_s=0$;  examples of latter class of zero-energy frozen state are states $|\psi(0)\rangle$ and $|\tilde \psi(0)\rangle$ with maximum bosonic dipole number discussed in the main text.  In contrast, Fock states with at least one rung having $|+\rangle$ or $|-\rangle$ states do not constitute eigenstates of $\hat{H}_F^{(1)}$. 

This analysis allows us to quantify the number of frozen states that appear in the system of length $L$. For a system with an odd $L$ at half-filling, at least one rung with either $|+\rangle$ or $|-\rangle$ will always be present. Consequently, there are no frozen states for odd $L$ at half-filling; however, a deviation from half-filling allows for frozen state for odd $L$. In contrast, for even $L$, half filled  Fock states with $L/2$ doubly occupied ($|\mathbb{1}\rangle$) rungs  and rest $L/2$ empty ($|{\mathbb 0}\rangle$) rungs may be constructed. These states do not have any rung with $|+\rangle$ and $|-\rangle$ and are therefore eigenstates of $\hat{H}_F^{(1)}$ at the special frequencies.  A chain of $L$ sites with $L/2$ of $|\mathbb{1}\rangle$'s and $L/2$ of $|\mathbb{0}\rangle$'s can be rearranged in $^LC_{L/2}$ number of ways leading to $N_{\rm frozen}= ^L C_{L/2}$. We note that the number of such frozen states grows exponentially with $L$ which is a hallmark of HSF.

{\it HSD of the largest fragment}: To ascertain the presence of 
strong HSF for $\hat{H}_F^{(1)}$  at special drive frequencies, we plot the ratio of the HSD of the largest fragment, $d_L$, to the total HSD, ${\mathcal D}$, as a function of $L$ for even(odd) $L$ in  Fig.~\ref{fig:large_H_dim_vs_L} (a) ((b)). In both cases, we find a linear decrease of $\ln(d_L/{\mathcal D})$; this indicates that the HSD of the largest sectors decreases exponentially $L$: $d_L/D \propto \text{exp}(-\alpha L)$ for $\alpha>0$. This exponential decay of $d_L$ serves as yet another indication of strong HSF.

\section{Chiral dynamics through $(N_d^{{\rm max}}-2)$ bosonic dipole sector}

The initial state we consider in the main text to study the chiral dynamics can be written as
\begin{equation}
    |\psi(0)\rangle =  \left| \begin{array}{cccccccccc}
1 & 0 & 1.~.~. & 1 & 0\\
1 & 0 & 1.~.~. & 1 & 0 
\end{array}\right\rangle 
=|\mathbb{1}\mathbb{0}\mathbb{1}.~.~.\mathbb{1}\mathbb{0}\rangle. \nonumber
\end{equation}
It's $Z_2$ partner state is given by $|\tilde \psi(0)\rangle=|\mathbb{0}\mathbb{1}\mathbb{0}.~.~.\mathbb{0}\mathbb{1}\rangle$. It is important to note that both of these are zero energy eigenstates of $\hat{H}_F^{(1)}$ at special drive frequencies. 
In Fig. 4 of the main text, we have shown that dynamics starting from initial state $|\psi(mT)\rangle$ proceeds through the states having $N_d= N_d^{\rm max}= N/4$ and $N_d^{\rm max}-2$ bosonic dipoles at large drive amplitudes and for special drive frequencies. The origin of such localization of the driven state in the Hilbert space can be traced back to the HSF of $\hat{H}_F^{(1)}$, which controls the dynamics up to a large prethermal timescale, 
at the special drive frequencies.

\begin{figure}[t!]
    \centering
    \includegraphics[width=1\columnwidth]{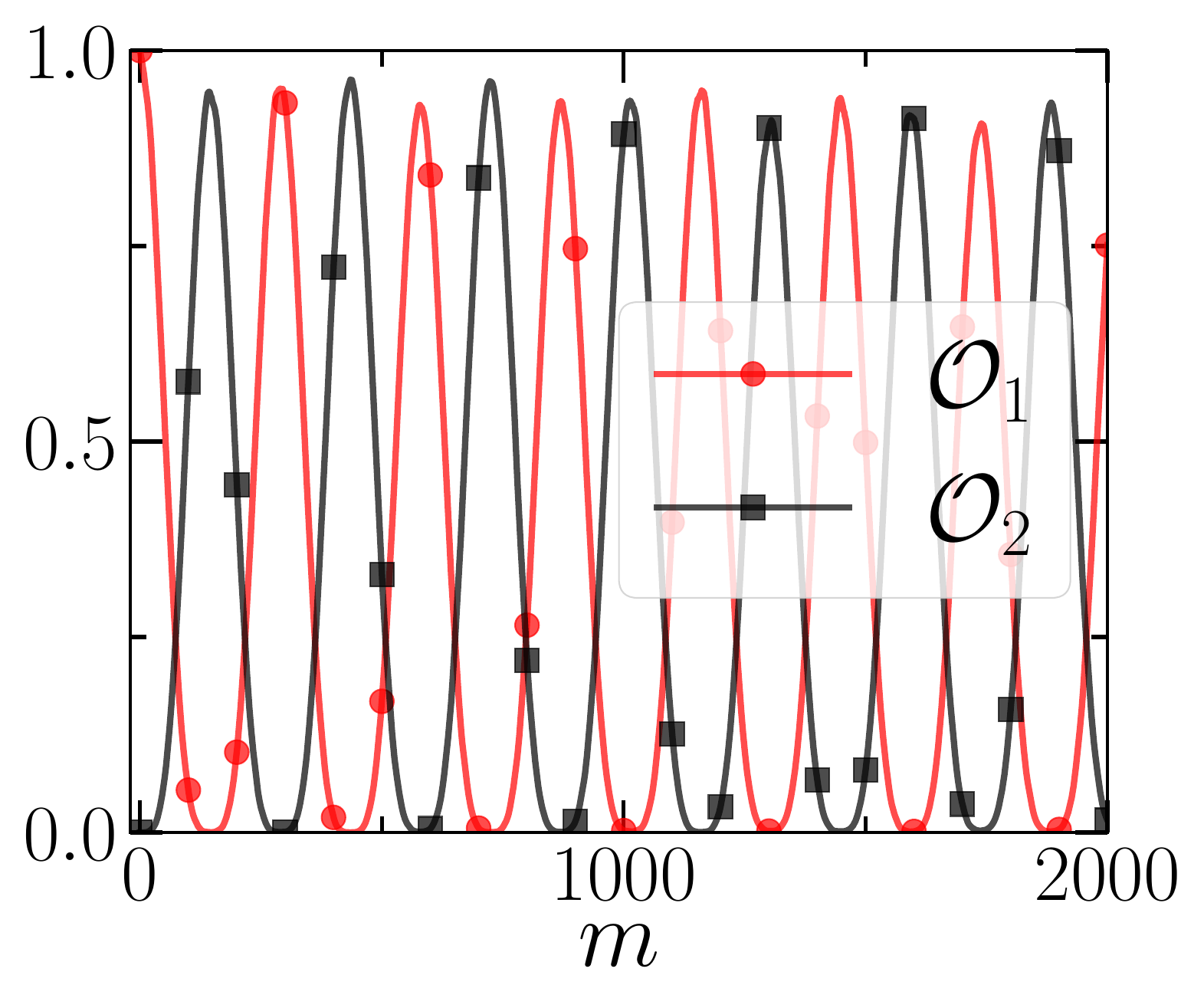}
    \caption{Plot of $\mathcal{O}_1 = |\langle|\psi(mT)|\psi(0)\rangle|^2$ (red circle and solid line) and $\mathcal{O}_2 = |\langle|\psi(mT)|\tilde \psi(0)\rangle|^2$ (black square and solid line) as a function of $m$ starting from initial state $|\psi(0)\rangle$. The calculations are done by setting $\omega_D=\omega_1^\ast$ and for $J_\perp/J = 1,~V_0/J =30,~V_\parallel/J=1$, and $\phi=0.75\pi$.}
    \label{fig:overlap_O1_O2}
\end{figure}

A schematic representation of such dynamics is shown in Fig.\ \ref{fig:toy_dynamics} for $L=10$. Our numerics, shown in Fig.\ \ref{fig:overlap_O1_O2}, finds that the dynamics of the driven state is well approximated by oscillation between $|\psi(0)\rangle$ and its $Z_2$ symmetric partner $|\tilde \psi(0)\rangle$. These oscillations are mediated by linear superposition of Fock states in the $N_d^{\rm max}-2$ bosonic dipole sector; the corresponding matrix elements connecting $|\psi(0)\rangle$ and $|\tilde \psi(0)\rangle$ occurs due to higher-order terms in $\hat{H}_F$. 

Such a dynamics constitutes a mechanism for restoration of $Z_2$ symmetry in a finite system. To see this, we consider a subspace of two states involving $|\psi(0)\rangle$ and $|\tilde \psi(0)\rangle$. Within this subspace, the effective Floquet Hamiltonian governing their dynamics can be understood as follows. Since both these states are zero energy states of $H_F^{(1)}$, the diagonal elements are close to zero; higher order diagonal terms do not lift the degeneracy between these two states due to the $Z_2$ symmetry and are therefore unimportant. The off-diagonal elements that lift the degeneracy of these states occur due to multiple hoppings originating from $H_F^{(2)}$ and $H_F^{(1)}$. The former controls hopping from $|\psi(0)\rangle$ to a state in the $N_d^{\rm max}-2$ bosonic dipole sector while the latter controls hopping within that sector. These processes are shown schematically in Fig.\ \ref{fig:toy_dynamics}. Thus in this two-state subspace, the eigenstates of $\hat{H}_F$ corresponds to bonding and anti-bonding combinations of $|\psi(0)\rangle$ and $|\tilde \psi(0)\rangle$ such that 
\begin{eqnarray} 
\hat{H}_F |\psi_{b(a)}\rangle = \epsilon_{b(a)}|\psi_{b(a)}\rangle.
\end{eqnarray}
Here $|\psi_{b(a)}\rangle = (|\psi(0)\rangle +(-) |\tilde \psi(0)\rangle)/\sqrt{2}$ \cite{flhsf1_1}. The energy splitting between these bonding and antibonding states $\Delta \epsilon = \epsilon_b-\epsilon_a$ is therefore small since its amplitude involves $O(L/2)$ hoppings at least two of which involve higher-order terms in $\hat{H}_F$. 

The oscillations of the overlap amplitude ${\mathcal O_{1,2}}$ can be understood qualitatively from the above considerations using an effective two-state description \cite{flhsf1_1}. Since $|\psi_{b(a)}\rangle$ are eigenstates of $\hat{H}_F$, starting from an initial  state $|\psi(0)\rangle$, the dynamics of the driven state can be thought as Rabi oscillations given by 
\begin{eqnarray} 
|\psi(t) \rangle &=& \cos (\Delta \epsilon t/(2\hbar)) |\psi(0)\rangle + \sin (\Delta \epsilon t/(2\hbar)) |\tilde \psi(0)\rangle. \nonumber\\
\end{eqnarray}
From this, we find 
\begin{eqnarray} 
{\mathcal O}_1 &=& |\langle|\psi(mT)|\psi(0)\rangle|^2 = \cos^2 (\Delta \epsilon t/(2\hbar)) \nonumber\\
{\mathcal O}_2 &=& |\langle|\psi(mT)|\tilde \psi(0)\rangle|^2= \sin^2 (\Delta \epsilon t/(2\hbar)).
\end{eqnarray}
We note that such an oscillation is a signature of both HSF which do not allow the drive state to spread in the Hilbert space and the $Z_2$ symmetry mentioned above. Therefore we do not expect such coherent oscillatory dynamics for other frozen Fock states for which no symmetric partner exists.

The many-body driven state $|\psi(mT)\rangle$ can therefore be written as a linear combination of states in the $N_d^{\rm max}$ and $N_d^{\rm max}-2$ dipole sector. Since $\hat J_c$ connects between these sectors (and also between different Fock states of the $N_d^{\rm max}-2$ bosonic dipole sector), we find a finite prethermal chiral current as discussed  in the main text. We note here that other Fock states which do not have a symmetric $Z_2$ partner can also yield finite chiral current due to localization fo the driven state originating from HSF. We have numerically checked, for a few such Fock states, that this is indeed the case; however, the magnitude of $\langle J_c\rangle_t$ is typically smaller for such states.

\end{document}